\documentclass[12pt]{jpconf}
\pdfoutput=1
\usepackage{graphicx}
\usepackage{footmisc}
\usepackage{color}
\usepackage{amsmath} 
\usepackage{array,relsize,float}
\usepackage[bitstream-charter]{mathdesign}

\renewcommand{\thefigure}{\arabic{figure}}

\def\beq{\begin{equation}}
\def\eeq{\end{equation}}
\def\bea{\begin{eqnarray}}
\def\eea{\end{eqnarray}}
\def\beqn{\begin{eqnarray}} 
\def\eeqn{\end{eqnarray}}

\def\bra#1{\langle{#1}|}
\def\ket#1{|{#1}\rangle}

\renewcommand{\thefootnote}{\fnsymbol{footnote}}

\begin{document}
\title{Geometry and causal flux in multi-loop Feynman diagrams}

\author{G. F. R. Sborlini$^{1,2}$\footnote[1]{\hspace{1mm}Speaker}}

\address{$^1$ Deutsches Elektronen-Synchrotron DESY, Platanenallee 6, 15738 Zeuthen, Germany.}
\address{$^2$ Instituto de F\'isica Corpuscular, Universitat de Val\`encia - Consejo Superior de Investigaciones Cient\'ificas, Parc Cient\'ific, E-46980 Paterna, Valencia, Spain.}

\ead{german.sborlini@desy.de}

\begin{abstract}
In this review, we discuss recent developments concerning efficient calculations of multi-loop multi-leg scattering amplitudes. Inspired by the remarkable properties of the Loop-Tree Duality (LTD), we explain how to reconstruct an integrand level representation of scattering amplitudes which only contains physical singularities. These so-called \emph{causal representations} can be derived from connected binary partitions of Feynman diagrams, properly entangled according to specific rules. We will focus on the detection of flux orientations which are compatible with causality, describing the implementation of a quantum algorithm to identify such configurations.
\end{abstract}

\setcounter{page}{1}

\section{Motivation and introduction}
\label{sec:intro}
The Standard Model (SM) is considered one of the most successful theories, since many predictions were found in complete agreement with the experiments. Recent technological advances are leading to a mayor improvement in the quality of the data collected from particle accelerators, unveiling the tiniest details of the structure of matter. The forthcoming upgrades of LHC and the planed future colliders \cite{Abada:2019lih,CEPCStudyGroup:2018ghi} will challenge the available theoretical results, forcing to achieve more precise predictions. 

In the context of high-energy physics, perturbation theory is widely applied to extract predictions from the models. Thus, higher orders are required in order to increase the accuracy of the theoretical results. This involves dealing with complicated multi-leg multi-loop Feynman diagrams and multi-particle phase-space integrals. To increase even more the difficulty, singularities and ill-defined expressions pop up in intermediate steps of the calculations. 

To tackle these problems, regularization techniques must be used. A frequent choice is Dimensional Regularization (DREG) \cite{Ashmore:1972uj,Cicuta:1972jf,tHooft:1972tcz}, which turns out to preserve most of the original symmetries but at the same time introduces issues related to the $D$-dimensional extension of the theory. Several regularization methods are available and novel ideas are being proposed to regularize the expressions keeping the original number of space-time dimensions~\cite{Gnendiger:2017pys,TorresBobadilla:2020ekr}. 

Regularizing the singularities only make them explicit, but several ingredients must be computed and added together in order to arrive to a finite physical result. By virtue of general theorems \cite{Kinoshita:1962ur,Lee:1964is}, virtual and real contribution must be simultaneously considered to ensure the cancellation of infrared (IR) divergences. The cancellation of additional IR singularities due to initial-state radiation (ISR) requires the introduction of proper counter-terms, which can be calculated within perturbation theory and involve the collinear splitting functions \cite{Altarelli:1977zs,Sborlini:2013jba,Sborlini:2014mpa,Sborlini:2014kla}. Also, the unrestricted loop integration in the virtual contributions leads to ultraviolet (UV) divergences, that can be removed through the renormalization procedure \cite{Collins:105730}.     

The purpose of this review is to explain novel technologies pointing towards a more efficient calculation of the loop contributions. In particular, these developments are done in the context of the Loop-Tree Duality (LTD) formalism \cite{Catani:2008xa,Rodrigo:2008fp,Bierenbaum:2010cy,Bierenbaum:2012th,deJesusAguilera-Verdugo:2021mvg}. The main advantage of this formalism is that Feynman loop integrals in Minkowski space are transformed into sums of phase-space integrals of tree-level like objects, i.e. the so-called \emph{dual terms}, naturally defined in Euclidean space. In this way, it is more transparent to establish a connection between the IR singularities in the real and the virtual contributions \cite{Buchta:2014dfa}, since they involve the same kind of integration variables. The LTD is the foundation of a disruptive method known as Four-Dimensional Unsubtraction (FDU) \cite{Hernandez-Pinto:2015ysa,Sborlini:2016gbr,Sborlini:2016hat,Driencourt-Mangin:2019aix}, that achieves a purely four-dimensional representation of physical observables by introducing proper kinematical mappings to combine the virtual and real contributions. In this way, the cancellation of IR singularities takes place directly at integrand level, before integration and thus by-passing any need of using DREG (or other regularization prescription).

The LTD approach turns out to be extremely useful to explore the causal structure of multi-loop multi-leg Feynman amplitudes. After presenting the basis of the LTD formulation, we will explain the deep connection between the geometrical interpretation of LTD and the causal structure of scattering amplitudes in Sec. \ref{sec:LTDintro}. In particular, we will briefly introduce some geometric rules to select configurations leading to causal entangled thresholds, in Sec. \ref{ssec:Selection}. Due to the fact that this involves testing the causality conditions on an increasingly large number of potential configurations, we review the basis of Grover's search algorithm in Sec. \ref{sec:GroverIntro}. Then, in Sec. \ref{sec:QAlgo}, we show a concrete proof-of-concept of a quantum algorithm to select those configurations compatible with causal momentum flow. Finally, in Sec. \ref{sec:conclusions}, we present the conclusions and give an outlook of interesting future research directions inspired by the reported results. 


\section{A brief introduction to Loop-Tree Duality}
\label{sec:LTDintro}
The LTD formulation is based on the application of Cauchy's residue theorem to multi-loop multi-leg Feynman integrals. It was first introduced at one-loop in Ref. \cite{Catani:2008xa}, where the connection to the classical Feynman Tree Theorem (FTT) was achieved through a suitable modification of the customary $\imath 0$ prescription. This led to the definition of the dual propagators which allows to re-write $N$-point one-loop integrals as sums of $N$ dual terms originated from single cuts \cite{Rodrigo:2008fp}. This approach was extended to deal with amplitudes at two-loops and beyond in Ref. \cite{Bierenbaum:2010cy}, and also to re-write amplitudes with higher-powers of the denominators \cite{Bierenbaum:2012th}. Besides that, LTD turns out to be particularly useful for exploring simplifications in asymptotic expansions at integrand level \cite{Driencourt-Mangin:2017gop,Plenter:2019jyj,Plenter:2020lop}, connections with the color-kinematics duality \cite{Jurado:2017xut} and efficient numerical integration strategies \cite{Buchta:2015wna,Buchta:2015xda}.

In the last years, a renewed interest on the topic attracted the attention of several groups and motivated new discoveries \cite{Aguilera-Verdugo:2019kbz,Driencourt-Mangin:2019yhu,Verdugo:2020kzh,Runkel:2019yrs,Runkel:2019zbm,Capatti:2019edf,Capatti:2019ypt,Capatti:2020ytd}. A novel formulation based on the calculation of nested residues \cite{Verdugo:2020kzh,JesusAguilera-Verdugo:2020fsn} allows to smoothly extend the LTD to multi-loop multi-leg amplitudes. By applying the Cauchy's residue theorem on the energy component, we can remove one degree of freedom per integral, thus reducing the number of integration variables and transforming the integration domain into an Euclidean space. Explicitly, given 
an $N$-point $L$-loop scattering amplitude in the Feynman representation, i.e.
\beqn
{\cal A}_F^{(L)} &=& \sum_l \, \int_{\ell_1 \ldots \ell_L} \, {\cal N}_l(\{q_i\},\{p_j\}) G_F(1,2,\ldots,n) \equiv \int_{\ell_1 \ldots \ell_L}  \,d{\cal A}_{N}^{\left(L\right)} \, ,
\label{eq:FEYNMAN}
\eeqn
we obtain the following integrand-level representation
\beqn
\nonumber d{\cal A}_D^{(L)}(1,\ldots,r;r+1,\ldots,n) &=& \sum_{i \in r} {\rm Res}\left(d{\cal A}_D^{\left(L\right)}{(1,\ldots,r-1;r,\ldots,n)}, {\rm Im}(q_{i,0})<0 \right) \, ,
\\ && \, 
\label{eq:IteracionR}
\eeqn
for the $r$-th step. The dual representation is obtained after $L$ iterations of Eq. (\ref{eq:IteracionR}), and the corresponding integration over the three-momenta of the $L$ loop variables. More details about the notation used here, as well as the associated technical details can be found in Refs. \cite{Verdugo:2020kzh,Aguilera-Verdugo:2020kzc,Sborlini:2021nqu}.

The reformulation of LTD based on nested residues presents some remarkable properties. In first place, after each iteration, there are some contributions that vanish. These are the so-called \emph{displaced poles}, and  they are associated to unphysical contributions originated in poles of the integrand with opposite residues. The formal proof of their cancellation was given in Ref. \cite{JesusAguilera-Verdugo:2020fsn}. 

\begin{figure}[ht]
\begin{center}
\includegraphics[width=0.65\textwidth]{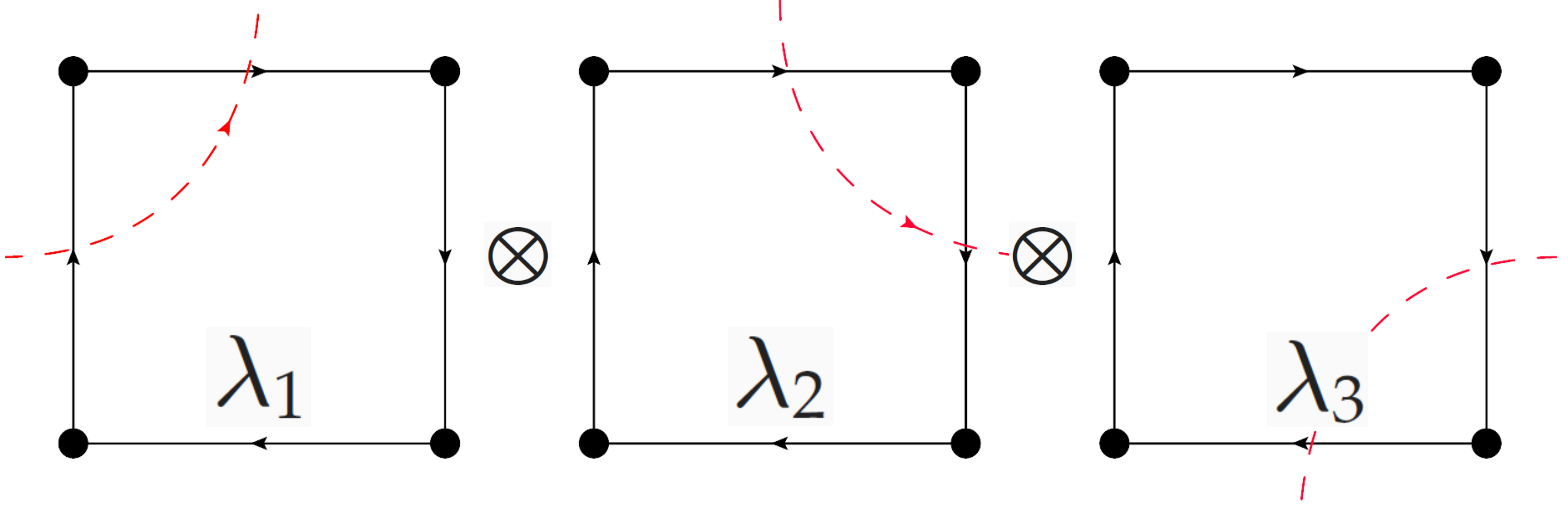} 
\caption{Compatible causal entangled threshold for a one-loop four-point function (i.e. a box). Since there are 4 vertices, $k=3$ thresholds can be simultaneously entangled. The configuration shown also exemplifies the geometrical compatibility rules described in Sec. \ref{ssec:Selection}.
\label{fig:0}}
\end{center}
\end{figure}

In second place, after the $L$-th iteration of Eq. (\ref{eq:IteracionR}) and adding all the terms, further cancellations take place. Only those contributions involving same-sign combinations of on-shell energies remain: these are the so-called \emph{causal contributions}. This behaviour is ultimately related to the deep connection between the LTD formalism and causality in quantum field theories \cite{Tomboulis:2017rvd,Runkel:2019yrs,Aguilera-Verdugo:2019kbz,Verdugo:2020kzh}. In fact, LTD naturally leads to a manifestly causal integrand-level representation of any multi-loop multi-leg scattering amplitude \cite{MANIFESTLYCAUSAL,Verdugo:2020kzh,TorresBobadilla:2021ivx,Sborlini:2021owe,Bobadilla:2021pvr}. After a careful exploration of several multi-loop multi-leg topologies, we obtain
\beqn
 {\cal A}_{N}^{\left(L\right)}&=& \int_{\vec{\ell}_{1},\cdots,\vec{\ell}_{L}}\, \sum_{\sigma \in \Sigma}  \frac{{\cal N}_{\sigma}(\{q_{r,0}^{(+)}\},\{p_{j,0}\})}{x_{n}} \,  \prod_{i=1}^{k} \frac{-1}{\lambda_{\sigma(i)}}\ + \ (\sigma \leftrightarrow \bar{\sigma})  \, ,
\label{eq:CAUSALREP}
\eeqn
which provides a causal representation in terms \emph{causal entangled thresholds}. The building blocks of this representation are the causal propagators, $\lambda_j^\pm  \equiv \sum q_{i,0}^{(+)} \pm k_{j,0}$, which involve positive on-shell energies and same-sign combinations of the energies of the external particles. Also, each causal propagator is in one-to-one correspondence with each possible physical threshold of the original amplitude. The number of thresholds (or, equivalently, causal propagators) that are simultaneously entangled determines the order of the diagram, i.e. $k$. The order is related to the number of off-shell lines in the traditional LTD representation, as well as the number of vertices $k=V-1$ \cite{Sborlini:2021owe}. A pictorial example of Eq. (\ref{eq:CAUSALREP}) for the one-loop four-point function is given in Fig. \ref{fig:0}. Concrete application examples and all-order generalizations of the LTD causal representation are given in Refs. \cite{Aguilera-Verdugo:2020kzc,Ramirez-Uribe:2020hes,TorresBobadilla:2021ivx,Bobadilla:2021pvr}.


\subsection{Geometry, causality and causal propagators}
\label{ssec:Geometry}
In the previous section, we presented two equivalent representations of the scattering amplitudes: the Feynman representation in Eq. (\ref{eq:FEYNMAN}) and the causal representation in Eq. (\ref{eq:CAUSALREP}). Whilst the first one is associated to Feynman graphs made of propagators, vertices and loops, the second one can be built from \emph{multi-edges}, vertices and \emph{eloops} \cite{Sborlini:2021owe,Ramirez-Uribe:2021ubp,Sborlini:2021nqu}. It turns out that the causal structure is codified in the so-called reduced Feynman graphs, which are built from the original diagram by merging all the propagators connecting the same vertices. These merged propagators are the multi-edges and their associated on-shell energy is given by $q_{G,0}^{(+)} = \sum_{i \in G} \, q_{i,0}^{(+)}$, i.e. the same-sign sum of all the on-shell energies of the lines connecting those vertices.

From the reduced Feynman graph, we can built the \emph{vertex matrix}, ${\cal V}$, which is equivalent to the adjacency matrix and contains information regarding how the different vertices are connected. It is possible to implement operations on ${\cal V}$ in order to identify the set of connected binary partitions of vertices, ${\cal P}_V^C$. Graphically, $p \in {\cal P}_V^C$ corresponds to cutting the graph into two connected diagrams. Moreover, we can introduce the concept of \emph{conjugated causal propagator}, $\bar{\lambda}_{p}$, by summing all the momenta flowing through the partition $p$ (or, equivalently, $p^c$). Since momentum conservation is enforced in each vertex, we have $\bar{\lambda}_{p}+\bar{\lambda}_{p^c}=0$ if the external particles fulfil momenta conservation. 

The motivation to provide such definitions is that the binary partitions are associated to thresholds of the diagram. According to Cutkosky rules \cite{Cutkosky:1960sp}, physical thresholds are originated by splitting or cutting the diagram in two pieces. Since causal propagators are in one-to-one correspondence to the physical thresholds, they must be also related to the binary connected partitions. In fact, we can build the set of causal propagators starting from $\bar{\lambda}_p$ and replacing the multi-edge energies by the corresponding on-shell energies. The presence of external momenta leads to a two-fold mapping to generate the casual propagators, i.e. there exists a transformation such that $\bar{\lambda}_p \to \lambda^{\pm}_p$ which depends on the orientation of the multi-edges. More details can be found in Ref. \cite{Sborlini:2021owe}.

To conclude this section, let us mention two extreme cases. Given a diagram with $V$ vertices, the possible number of multi-edges fulfills $V \leq M \leq V(V-1)/2$. A graph with $M=V$ is minimally connected, whilst those with $M=V(V-1)/2$ are maximally connected. In Refs. \cite{TorresBobadilla:2021ivx,Bobadilla:2021pvr}, maximally connected graphs (MCG) are the basic building blocks of an all-order causal representation. The advantage MCG is that all the binary partitions are connected (${\cal P}_V^C={\cal P}_V$), which implies that the number of causal propagators is maximal. On the opposite side, minimally connected graphs (mCG) are always one eloop diagrams with the minimal number of causal propagators. We will do some further comments on these cases in the next sections.


\subsection{Selection rules}
\label{ssec:Selection}
Once the causal propagators are identified, we need to introduce a set of geometrical rules to build all the possible compatible causal entangled thresholds, $\bar{\Sigma}$. Given a topology with $V$ vertices and $M$ multi-edges, we will need to entangle $k=V-1$ causal thresholds. Then, we impose the following criteria:
\begin{enumerate}
 \item \emph{No uncut multi-edges}: If we overlap all the causal thresholds (as the ones shown in Fig. \ref{fig:0}), all the internal multi-edges must be crossed by the causal cuts.
 \item \emph{Disjoint separation}: If we represent each causal propagator with a line cutting the multi-edges, then these lines can not intersect each other.
 \item \emph{Consistent causal flow}: The multi-edges associated to a partition must be aligned, and flow to different partitions.
\end{enumerate}
It is worth appreciating that, for some topologies, there will be degenerated contributions to Eq. (\ref{eq:CAUSALREP}). In order to obtain a non-degenerated subset, $\Sigma$, additional steps are required, as carefully explained Ref. \cite{Sborlini:2021owe}. The last criteria involves studying all the possible orientations of multi-edges, i.e. $2^M$ configurations. The fact that the aligned momenta must go from one partition to a different one implies that there can not be loops or cycles. Thus, the situation reduces to the identification  of \emph{directed acyclic graphs}. Once all the causal orientations are properly selected, we can dress the graphs with the entangled thresholds selected by criteria 1-2, and generate the set $\bar{\Sigma}$.


\begin{figure}[b]
\begin{center}
\includegraphics[width=0.25\textwidth]{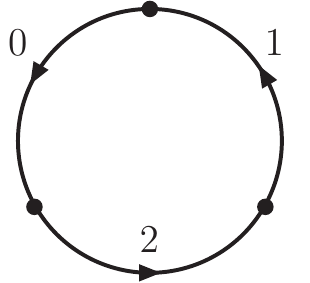} \quad
\includegraphics[width=0.65\textwidth]{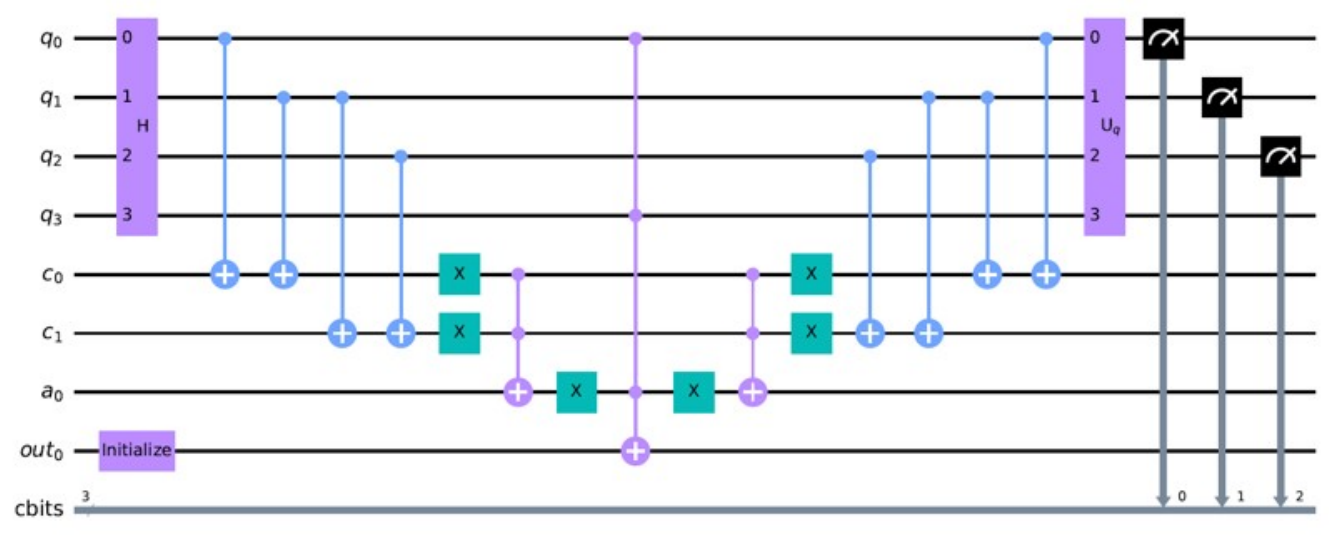} 
\caption{(Left) One eloop three-vertex topology with the corresponding labelling of the multi-edges. (Right) Graphical representation of the quantum circuit implemented in \texttt{Qiskit}. The different logical gates are drawn, as well as the input ($\{q_i\}$), the auxiliary ($\{c_j\}$ and $\{a_j\}$) and output ($\ket{out_0}$) registries.  
\label{fig:1}}
\end{center}
\end{figure}

\section{Quantum computing and Grover's algorithm}
\label{sec:GroverIntro}
Testing the causality conditions described in the previous section is a very time-consuming task. When the complexity of the diagram increases, i.e. by adding more vertices $V$ or multi-edges $M$, the number of combinations grows very fast. However, the criteria can be executed in parallel and enhance the performance of the computation.

As mentioned in Sec. \ref{ssec:Selection}, we can start by looking for all the possible directed acyclic graphs. Even if there are several well-performing algorithms \cite{Even20111} for classical computers, the complexity of the problem scales very fast with the number of multi-edges. Thus, in the search of more efficient strategies, we tried a novel approach based on quantum algorithms in Ref. \cite{Ramirez-Uribe:2021ubp}. In particular, we relied on Grover's search algorithm \cite{10.1145/237814.237866,Bennett:1996iu,Grover:1997ch} to test the causal flow conditions simultaneously for all the configurations. 

To explain the basis of the algorithm, let us consider the space $\Omega$ composed by $N$ states. A sub-space $\omega$ with dimension $r$ identifies the \emph{winning} states, i.e. the ones that we want to select. Then, we can define the two states
\beq
\ket{q}= \frac{1}{\sqrt{N}} \sum_{\alpha \in \Omega} \ket{\alpha} \, , \quad \ket{w}= \frac{1}{\sqrt{r}} \sum_{\alpha \in \omega} \ket{\alpha} \, ,
\label{eq:StateDefinition}
\eeq
where $\ket{q}$ ($\ket{w}$) corresponds to the uniform superposition of all the (winning) states. We will consider $\ket{q}$ as the initial state for running the algorithm, and introduce
\beq
U_w= {\rm Id} - 2\ket{w}\bra{w} \, , \quad U_q = 2 \ket{q}\bra{q} - {\rm Id} \, ,
\label{eq:OperatorDefinition}
\eeq
which are the so-called oracle and diffusion operators, respectively. Notice that $U_w \ket{x} = - \ket{x}$ if $\ket{x} \in \omega$ and leaves the state unchanged if it belongs to the orthogonal space $\omega^c$ (i.e. the space of non-winning states). Grover's querying algorithm relies on the repeated application of these operators, which leads to
\beq 
\ket{q'} \equiv (U_w \, U_q)^n \ket{q} = \sin\left((2n+1)\beta\right) \ket{w}+ \cos\left((2n+1)\beta\right) \ket{w^c}\, ,
\eeq
after $n$ iterations and $\beta=\arcsin(r/N)$ is the original projection of $\ket{q}$ over the winning space. In this way, if $r/N$ is small enough, it is possible to amplify the probability of the winning states w.r.t. those belonging to the orthogonal space after a few iterations of the algorithm.


\section{Quantum algorithm for causal flux identification}
\label{sec:QAlgo}
Finally, we explain how to profit from the quantum parallelization to efficiently identify the casual flux configurations. In first place, we associate a qubit to each multi-edge: $\ket{q_i}=\ket{1}$ if the flow agrees with the initial choice and $\ket{q_i}=\ket{0}$ if it is reversed. In this way, our quantum circuit will have the qubits $\{q_0,\ldots,q_{M-1}\}$ to describe the flow configuration. The uniform superposition state $\ket{q}$ is achieved through the application of Hadamard operators.

The explicit implementation of the oracle and diffusion operators requires to use logical gates and introduce auxiliary qubits. This is because qubits are quantum states: they can not be erased and re-used in the same way that we do with classical bits. Thus, we compare the fluxes of the different adjacent multi-edges and store the output in the auxiliary registers $\{c_i\}$. Using this information, we introduce another logical clause to probe whether the multi-edges (codified in the qubits $\{q_i\}$) are part of a cycle or not: the result is stored in the registry $\{a_j\}$. With all these elements, we build the marker function $f(a,q)$ and define Grover's oracle.

\begin{figure}[b]
\begin{center}
\includegraphics[width=0.8\textwidth]{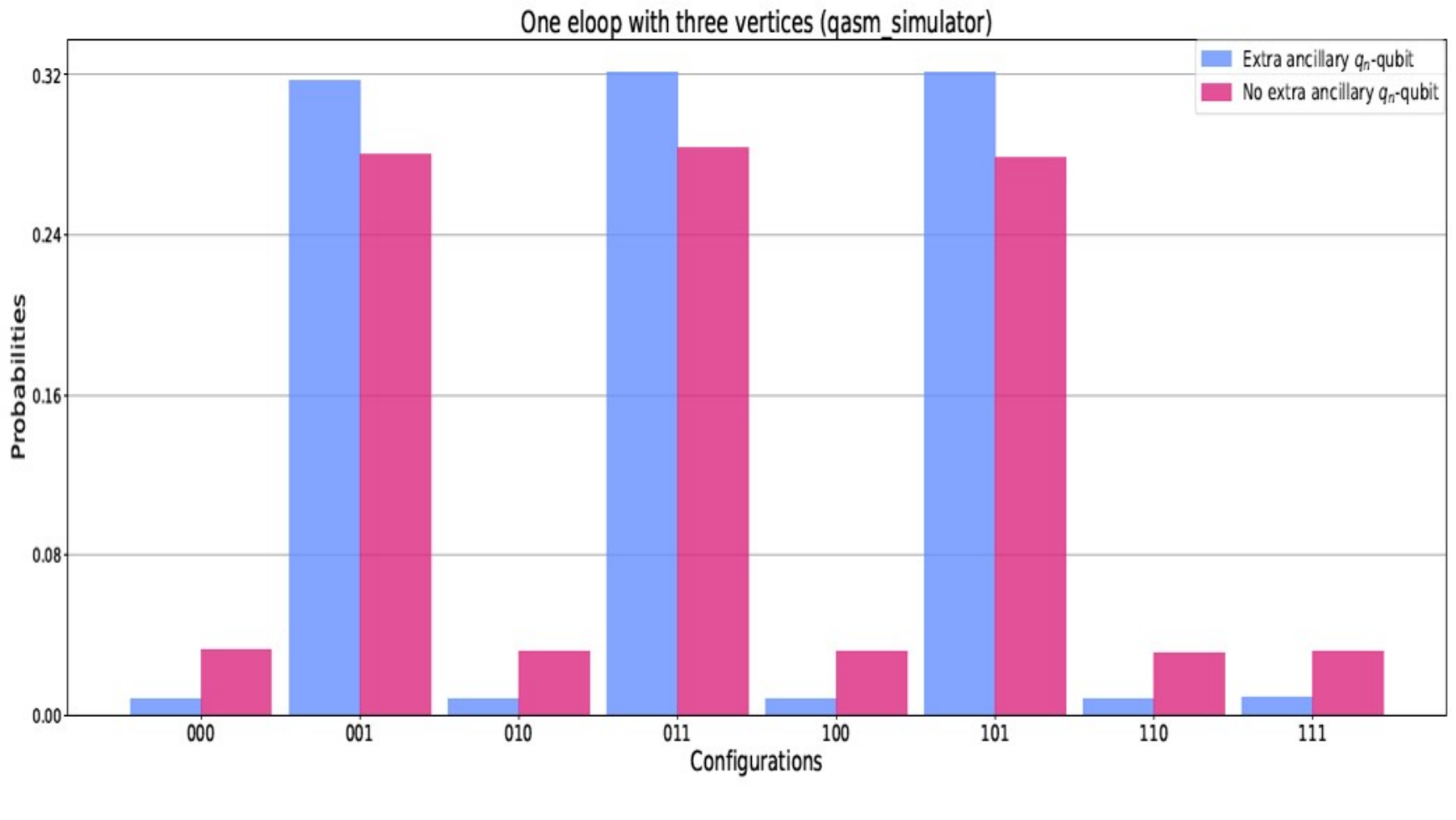} 
\caption{Probability distribution of the different configurations for an one eloop topology with three vertices. We appreciate that the algorithm successfully enhanced the probability of the three states satisfying the casual flow condition.
\label{fig:2}}
\end{center}
\end{figure}

As a practical example, we consider the identification of all the causal-compatible flux configurations for a three-vertex one-eloop topology (i.e. a \emph{triangle}). The circuit, as implemented in \texttt{Qiskit}, is shown in Fig. \ref{fig:1} (right). Since the number of solutions is roughly one half of the total states, we include an additional qubit to achieve the amplification of the winning states \cite{Ramirez-Uribe:2021ubp}. With this modification, we successfully identify the 3 causal-compatible configurations. The corresponding probability distributions are shown in Fig. \ref{fig:2}, where we can clearly appreciate that $\{\ket{001},\ket{011},\ket{101}\}$ describe directed acyclic graphs following the convention of Fig. \ref{fig:1} (left).

In general, the number of causal-compatible solutions depends on the topological complexity of the diagram (i.e. the number of vertices and eloops). For mCG diagrams, the presence of only one eloop leads to a weak constraining of the orientations: in this case, the ratio of winning over total states ($r/N$) tends to 1 as the number of vertices increases. On the other hand, MCG are extremely constrained systems and $r/N$ goes rapidly to 0.

\section{Conclusions and outlook}
\label{sec:conclusions}
The causal structure of multi-loop multi-leg scattering amplitudes contains all the relevant information to characterize their physical singularities. In combination with the Loop-Tree Duality formalism \cite{Verdugo:2020kzh}, causal representations lead to a more efficient numerical integration because they only contain physical singularities: instabilities due to spurious divergences are avoided. Moreover, the integration domain is transformed from a Minkowski to an Euclidean space, thus paving the way for a natural integrand-level combination with the real-radiation contributions.

Recently, there have been several developments to reconstruct dual causal representations by-passing the direct calculation of nested residues \cite{Bobadilla:2021pvr,Sborlini:2021owe}. One of these strategies relies on geometrical algorithms to detect all the possible causal propagators and identify the set of compatible causal entangled thresholds. In order to do that, we start from the reduced Feynman graph and associate a causal propagator to each possible connected binary partition of vertices. If the graph has $V$ vertices, we entangle $k=V-1$ causal thresholds in such a way that they fulfill specific selection criteria.

One of these selection criteria is related with the causal-compatible flux, which turns out to be equivalent to the identification of acyclic graphs \cite{Ramirez-Uribe:2021ubp}. In this review, we elaborate on the implementation of a quantum algorithm, based on Grover's methods, to efficiently detect acyclic directed graphs. It constitutes the first proof-of-concept of such an algorithm for bootstrapping the causal representation of Feynman scattering amplitudes, and we illustrate its application with a one eloop three-vertex topology.

It is interesting to notice that the performance of the algorithm increases as the ratio of casual versus total configurations decreases, which occurs as the topologies become more complicated (i.e. more vertices and more multi-edges). Moreover, since the total number of compatible causal entangled thresholds is small compared to the total possible combinations, we expect that a variation of this algorithm could efficiently reconstruct the whole causal representation. This supports the claim that the irruption of quantum technologies could allow to overcome many of the current bottlenecks in theoretical high-energy physics.

\section{Acknowledgments}
I would like to thank G. Rodrigo, S. Ram\'irez-Uribe, A. Renter\'ia-Olivo (IFIC), R. Hern\'andez-Pinto (UAS) and P. Marquard (DESY) for fruitful discussions and comments on the manuscript. This article is based upon work from COST Action PARTICLEFACE CA16201, supported by COST (European Cooperation in Science and Technology).

\section{References}

\providecommand{\href}[2]{#2}\begingroup\raggedright\endgroup

\end{document}